\documentclass[12pt,preprint]{aastex} 

\def \chandra{{\it Chandra}}
\def \hrci{{HRC-I}}
\def \ctrt{{ct~s$^{-1}$}}
\def \rfrac{{\rho_{{\rm frac}}}}
\def \rdiff{{\rho_{{\rm diff}}}}
\def \rchis{{\rho_{\chi^2}}}
\def \rt{{R}}
\def \lc{{C}}
\def \hz43{{HZ\,43}}
\def \dt{{\delta}}
\def \tscale{{\tau}}

\shorttitle{Capella Variability}
\shortauthors{Kashyap \& Posson-Brown}

\begin{document}

\title{Short Timescale Coronal Variability in Capella}

\author{
Vinay L.\ Kashyap\altaffilmark{1}, 
Jennifer Posson-Brown\altaffilmark{1}
}
\altaffiltext{1}{Harvard-Smithsonian Center for Astrophysics, 60 Garden St., Cambridge, MA 02138}

\begin{abstract}
We analyze 205~ks of imaging data of the active binary, Capella,
obtained with the \chandra\ High Resolution Camera Imager (HRC-I)
to determine
whether Capella shows any variability at timescales $<50$~ks.  We find
that a clear signal for variability is present for timescales
$\lesssim20$~ks, and that the light curves show evidence for excess
fluctuation over that expected from a purely Poisson process.  This
overdispersion is consistent with variability at the 2-7\% level, and
suggests that the coronae on the binary components of Capella are
composed of low-density plasma and low-lying loops.
\end{abstract}

\keywords{ stars: activity, stars: coronae, X-rays: stars }

\section{Introduction}

Capella ($\alpha$~Aur; G1\,III/G8\,III) is the strongest non-solar
coronal source accessible to high-sensitivity high-energy
telescopes, and has been a common calibration target for
X-ray and EUV instruments such as {\sl EUVE}, \chandra,
XMM-{\sl Newton}, etc.
It is a remarkably stable source, with no discernible flaring
activity.  Even though the emission structure has shown changes,
especially in the high-temperature regime (see e.g., Dupree
et al.\ 1996, Young et al.\ 2001), and there is considerable
evidence for the dominant emission to change between the G1\,III
primary and the G8\,III secondary (e.g., Linsky et al.\ 1998,
Johnson et al.\ 2002, Ishibashi et al.\ 2006), the overall
luminosity has remained steady over many years.
For instance, Argiroffi et al.\ (2003) detected a change of
3\% in \chandra\ HRC-S/LETGS data over the span of a year, and found
no variability at timescales of $0.1-10$~ks.  Recently, analysis
of \chandra\ ACIS-S/HETGS data (Raasen et al.\ 2007, Westbrook et al.\ 2007)
found variations over long timescales, such as an $\approx20\%$
intensity enhancement in early 2006, but no evidence for any
variability at timescales $<50-100$~ks.

Variability in stellar coronae is ubiquitous, and has been detected
in all types of coronally active stars
(Stassun et al.\ 2006,
Caramazza et al.\ 2007,
Westbrook et al.\ 2007,
G\"{u}del 2004, and references therein)
and over a wide range of timescales
(Kashyap \& Drake 1999,
Favata et al.\ 2005,
Stassun et al.\ 2006,
Pease et al.\ 2006,
Colombo et al.\ 2007).	
This variability can occur due to many causes, ranging from
cyclical dynamo variations (timescales of decades to years), to
rotational modulation (months to hours), to
active region evolution (hours to days), to
flaring (hours to minutes).
Generally, active stars are characterized by recurrent flares
(see e.g., G\"{u}del 2004)
that are recognized in X-ray and EUV light curves as sudden
increases in the luminosity followed by a slower decay.  However,
as activity increases, the flares start to occur closer in time
to each other, and it becomes increasingly difficult to resolve
them in the light curve (cf.\ Kashyap et al.\ 2001).  Prominent
flares are nevertheless detected in numerous active binaries
(see e.g., Osten \& Brown 1999, Osten et al.\ 2004).

Despite being one of the more coronally active stars, with a
strong high temeperature emission component (Brickhouse et al.\ 2000),
flares have never been observed on Capella (Table~\ref{t:capella}).
It is unknown whether this is due to a lack of flaring activity
to contribute to the heating, or due to a preponderance of flares
such that individual events cannot be distinguished (cf.\ Kashyap
et al.\ 2001).  Here we consider recent observations of Capella
made with the \chandra/\hrci\ (\S\ref{s:data}).  We analyze these
data and find that variability indeed can be detected at short
timescales (\S\ref{s:analysis}), suggesting that the latter
explanation is more plausible.

\begin{table}[htb!]
\begin{center}
\caption{Capella stellar properties \label{t:capella}}
\medskip
\begin{tabular}{rl}
\hline\hline
\multicolumn{2}{l}{$\alpha$~Aur / GJ 194 / HD 34029 / HIP 24608 / HR 1708} \\
\hline
(R.A., Dec)$_{ICRS 2000.0}$ & (05:16:41.3591, +45:59:52.768) \\
distance & 13.4 pc \\
orbital period & 104 days \\
Components & G1\,III / G8\,III \\
Separation & 109 R$_\odot$ \\
Mass & 2.56 / 2.69 [M$_{\odot}$] \\
Radius & 9.2 / 12.2 [R$_{\odot}$] \\
M$_{\rm V}$ & 0.14 / 0.25 \\
$B-V$ & 0.74 / 0.87 \\
rotational velocity & 36 / 3 [km~s$^{-1}$] \\
\hline
\end{tabular}
\end{center}
\end{table}


\section{Data \label{s:data}}

\begin{figure}
\centerline{\includegraphics[width=5.5in,angle=90]{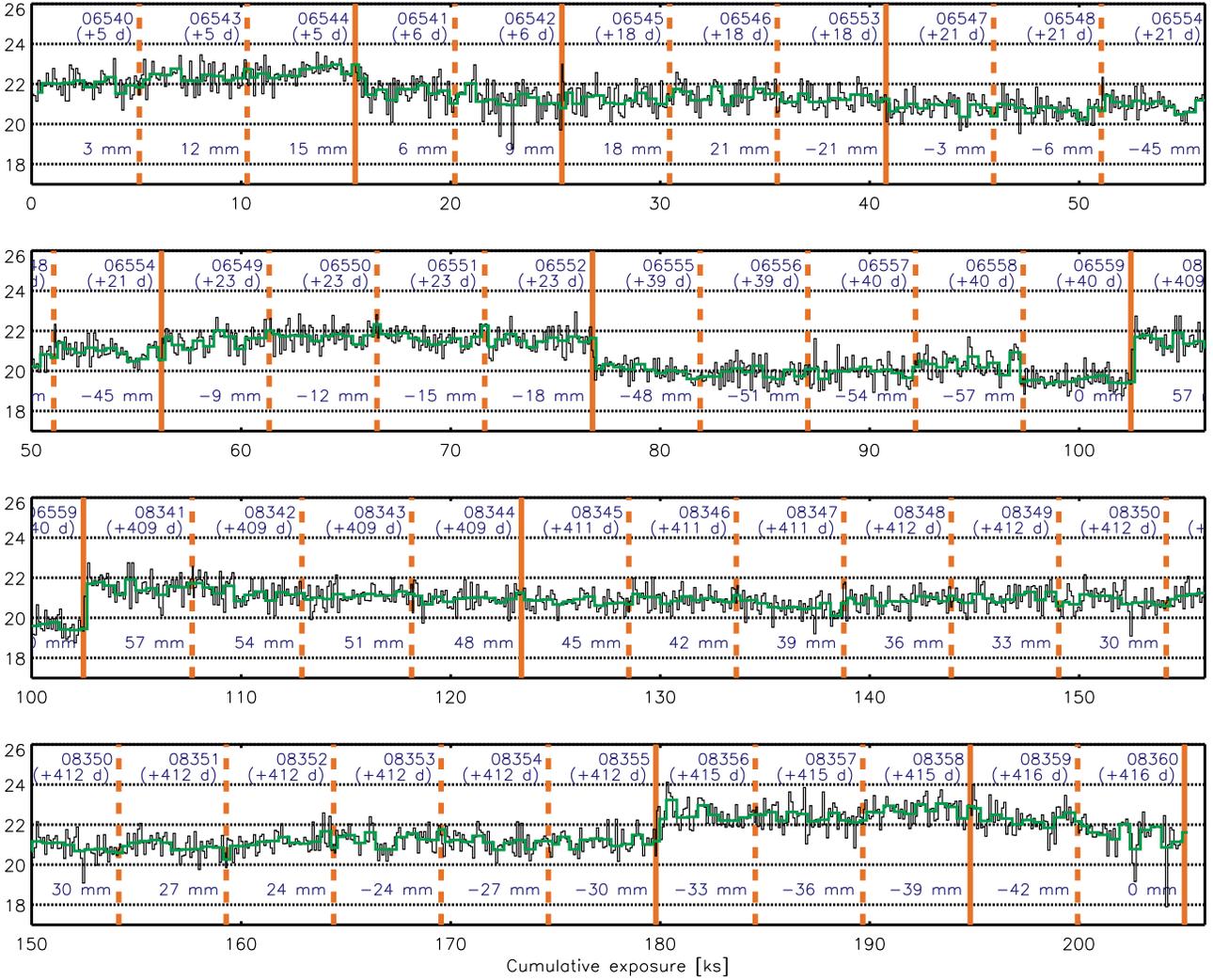}}	
\caption{The combined light curve of all the \chandra/\hrci\ observations
of Capella, spanning 205~ks.  The black histogram denotes the count rate
for a binning of 100~s, and overlaid on it is the count rate for a binning
of 500~s (green histogram).  The data gaps between observations are excluded,
and indicated by vertical red lines (solid when the gaps are $>100$~s,
dashed otherwise).  The data comprise 40 ObsIDs (noted at the top of
each segment, along with the day since 2005-dec-01 that the observation
started).  The SIM offset at which each observation is carried out is
indicated at the bottom of each segment.
\label{f:ctrt}}
\end{figure}

Capella was observed as a calibration target with the \chandra/\hrci\ over
two cycles from December 2005 to January 2007 (see Figure~\ref{f:ctrt}).
All the observations were carried out at the telescope aimpoint, but at
different locations on the detector corresponding to different offset values
of the Science Instrument Module (SIM).
The count rates shown in Figure~\ref{f:ctrt} have been corrected
for the QE (quantum efficiency) values at the observation location
and thus represent flat-fielded light curves matched to the QE
at the nominal aimpoint.  The QE corrections are made separately
for each bin of the light curves as the source dithers across the
detector.  Note however that this detailed correction is ignorable;
it causes changes of $<<1\%$ when compared with count rates corrected
with a QE averaged over the entire dither pattern (see also \S\ref{s:hz43}).
We have reduced the data using the \chandra\ software for the interactive
analysis of observations (CIAO v3.4) and using the most recent calibration
products (CALDB v3.2).
The high count rates observed ($\gtrsim20$~\ctrt), coupled
with the sharp point spread function (PSF; it falls to 1\% of
the maximum at $\approx1.3''$ away from the peak), provides an unprecedented
opportunity to study small changes in the X-ray brightness of Capella.
The source counts are extracted from a circle with radius 8x the size
of the PSF ($\approx10''$), and the background is locally estimated from
a surrounding annulus of radii ($\approx10-33''$).
Assuming a Raymond-Smith thermal emission model with a dominant
temperature component at 6~MK and a H column N$_H=10^{18}$~cm$^{-2}$,
we find with WebPIMMS\footnote{
{\tt http://asc.harvard.edu/toolkit/pimms.jsp}
}
that the counts-to-energy conversion factor is
$\sim~5.7-6.0\times10^{-12}$~ergs~cm$^{-2}$~ct$^{-1}$
for different metallicities.  This suggests X-ray luminosities of Capella
in the 0.15-4.5~keV passband of $2.4-2.9\times10^{30}$~ergs~s$^{-1}$
(cf.\ L$_X=3.8\times10^{30}$~ergs~s$^{-1}$ based on {\sl Einstein}/IPC
observations; 
Strassmeier et al.\ 1993).

\section{Analysis and Discussion \label{s:analysis}}

\subsection{Autocorrelation \label{s:autocorr}}

\begin{figure}
\plotone{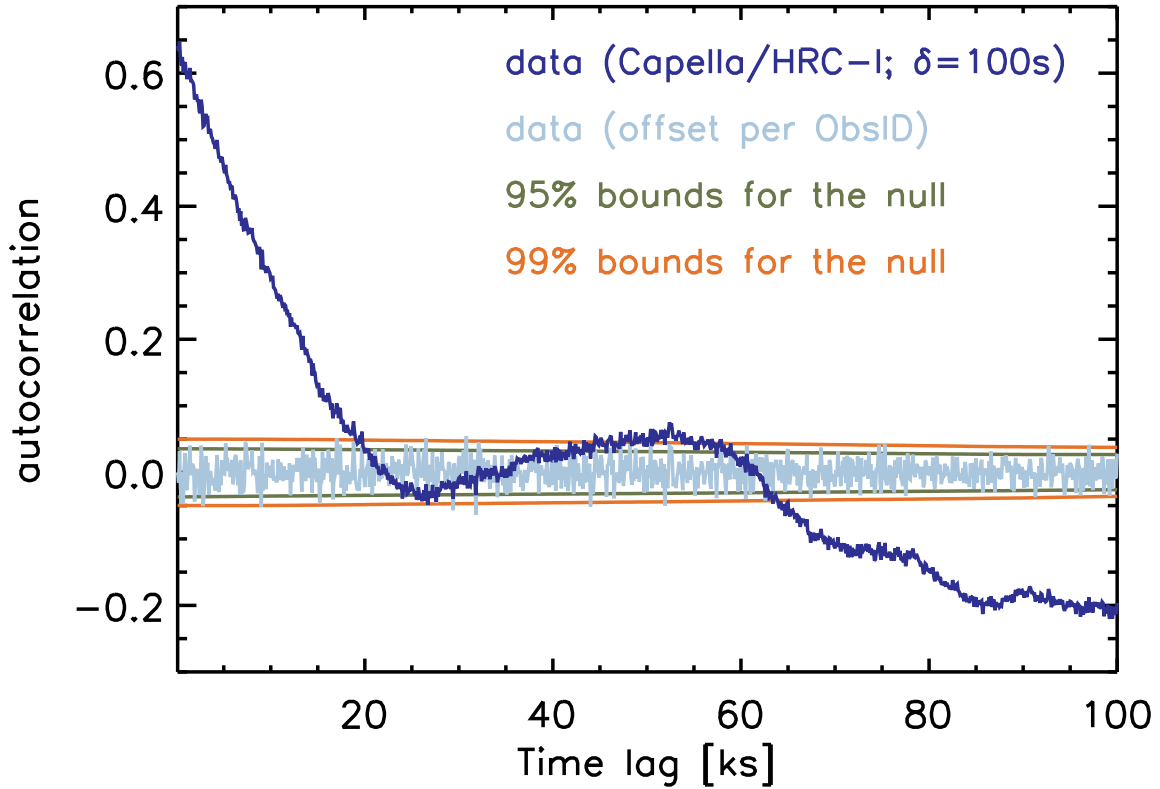}	
\caption{Autocorrelation in the light curve of Capella.  The autocorrelation
for a light curve binned by 100~s is shown as the dark blue curve, computed
for different lag times.  The expected autocorrelation for a light curve
with no variability is 0 for all non-zero lag times, and the 95\% and 99\%
uncertainty bounds on it are shown as horizontal lines (green and red
respectively).  The bounds are computed via Monte Carlo simulations of
a flat light curve with the same mean and number of bins.  There is
clear evidence for variability at timescales $\lesssim20$~ks.  In contrast,
the autocorrelation for a light curve constructed by offsetting it by the
mean count rate in each ObsID (pale blue curve) shows no evidence of
variability.
\label{f:autocorr}}
\end{figure}

It is clear from the \hrci\ light curve (Figure~\ref{f:ctrt}) that
Capella undergoes slow changes in its luminosity over timescales of
weeks and months, with count rates ranging from $\approx20-23$~\ctrt.
For instance, note the drop in intensity between
ObsIDs 6552 and 6555, which are separated by 16~days,\footnote{
The SIM offset also changes from $-18$ to $-48$~mm between these
two observations, and it could be argued that uncalibrated differences
in the QE uniformity may account for the drop in intensity.  However,
the similarity of count rates between observations 6558 and 6559,
both of which were carried out on the same day, but at SIM offsets
of $-57$ and $0$~mm respectively, indicates that the drop in intensity
is real.
}
and the difference between ObsIDs 6559 and 8360, which are done
$\approx1$~yr apart, but are both carried out at the same detector
location (thus precluding calibration differences as a factor; see
\S\ref{s:hz43} below).

This impression is confirmed by an autocorrelation analysis;
we construct count rate light curves $\rt(t_i;\dt)$ at various
binning sizes $\dt$ and compute the autocorrelation
$$ 
P(t_k;\dt) = \frac{\sum_{i=1}^{N_\dt-k} (\rt(t_i;\dt) - \nu(\dt))(\rt(t_{i+k};\dt)-\nu(\dt)) }{ \sum_{i=1}^{N_\dt} (\rt(t_i;\dt)-\nu(\dt))^2} \,,
$$ 
where
$
\nu(\dt) 
$
is the average
count rate in the $N_\dt$ bins in the light curve.  The autocorrelation for
$\dt=100$~s is shown in Figure~\ref{f:autocorr} as the dark blue curve
(curves for other bin sizes are similar), along with estimates of the
95\% and 99\% uncertainties for each lag time $t_k$.  These
uncertainties are computed via Monte carlo simulations, by constructing
1000 light curves with the same number of bins as in $\rt(t_i;\dt)$, as
Poisson deviates for an unvarying source intensity of $\nu(\dt)\cdot\dt$.
The autocorrelation drops linearly until it becomes indistinguishable from
statistical noise at a lag time of $\approx20$~ks; this is the typical
signature of variability which occurs at timescales $\lesssim20$~ks, 
such that count rates that are separated by longer timescales are
essentially uncorrelated.
At larger lag times, $\sim30-60$~ks, the autocorrelation appears to rise
again, but this is not distinguishable from statistical noise.\footnote{
At even larger lag times, $t_k>65$~ks, the autocorrelation drops further
and becomes significantly $<0$, suggesting that the count rates are
anticorrelated at large timescales, i.e., the intensity tends to 
fluctuate over long temporal separations.
However, these values are not physically
meaningful, since the data gaps between observations are large and the
lag time ceases to be a useful construct.  Note that typical observation
times are 5~ks, and the observation times for contiguous ObsIDs ranges
from 12 to 62~ks, with a median of 19~ks (see Figure~\ref{f:ctrt}).
}

This is consistent with the results found by Raasen et al.\ (2007) and
Westbrook et al.\ (2007), who found similar variations over similar
timescales.  However, at a counts intensity level of $\approx2$~\ctrt\
(obtained from ACIS-S/HETG dispersed counts) they were unable to detect
any variability at timescales $\lesssim50$~ks corresponding to the
durations of the observations, 
even using sophisticated algorithms such as the one described by
Gregory \& Loredo (1992).
Here, we observe the source with a counts intensity an order of magnitude
higher, and are thus able to investigate the variability at shorter
timescales.  Note that Argiroffi et al.\ (2003) also find no variability
at timescales $<10$~ks, but again, the HRC-S/LETGS data they rely on
has count rates of $\lesssim3$~\ctrt, which is too small to detect the
existence of variability on Capella.

Because the \hrci\ observations are done in short segments, it is useful to
consider the effect of removing large timescale effects on the
autocorrelation.  We thus reconstruct the light curve by offsetting
that computed in each ObsID by the average intensity of the source
during that ObsID (i.e., for each segment the average count rate is
set to $0$), and recompute the autocorrelation.  This has the
effect of completely removing variations at timescales greater
than $\approx\frac{1}{2}$ of the typical exposure time, and will reveal
any variability that may exist at very small timescales.  The result
of this is shown as the pale blue curve in Figure~\ref{f:autocorr}.
This is everywhere consistent with no variability.  Because the typical
observation time is 5~ks, this suggests that there is no variability
variability on Capella at timescales $\lesssim3$~ks.  Note that offsetting
the light curves by different amounts at different times, as we have
done in this exercise, introduces additional statistical uncertainty
into the results because of a non-stationary bias, and therefore
the power of the test to detect a variability signal is decreased.
In order to determine whether there does exist variability over
timescales of $\sim5$~ks, we test the dispersion of the fluctuations
(see \S\ref{s:overdisp}).

\subsection{Overdispersion \label{s:overdisp}}


The large count rates of Capella seen with the \hrci\ afford
us the capability to analyze the light curve at short timescales.
In order to test the constancy of the intensity within each ObsID,
we carry out a Kolmogorov-Smirnoff (K-S) test on the photon arrival time
data in each ObsID.  A large fraction of the datasets show evidence
that the null hypothesis of no variation within an observation can be
rejected: 17\% have $p<0.05$, and 37\% have $p<0.1$.  This is not
strong evidence for short timescale variability, but the repeated 
rejection of the null at a frequency larger than expected indicates
the possible existence of intermittent variability.  Also note that
the K-S test generally has low power in detecting small, slow
variations, and a better test is required.  For that, we consider
whether the observed fluctuations in the light curve are consistent
with statistical deviations.  Such a test is also of value in
establishing the magnitude of the residual calibration errors (see
\S\ref{s:hz43}).

\begin{figure}
\centerline{\includegraphics[width=5.5in,angle=90]{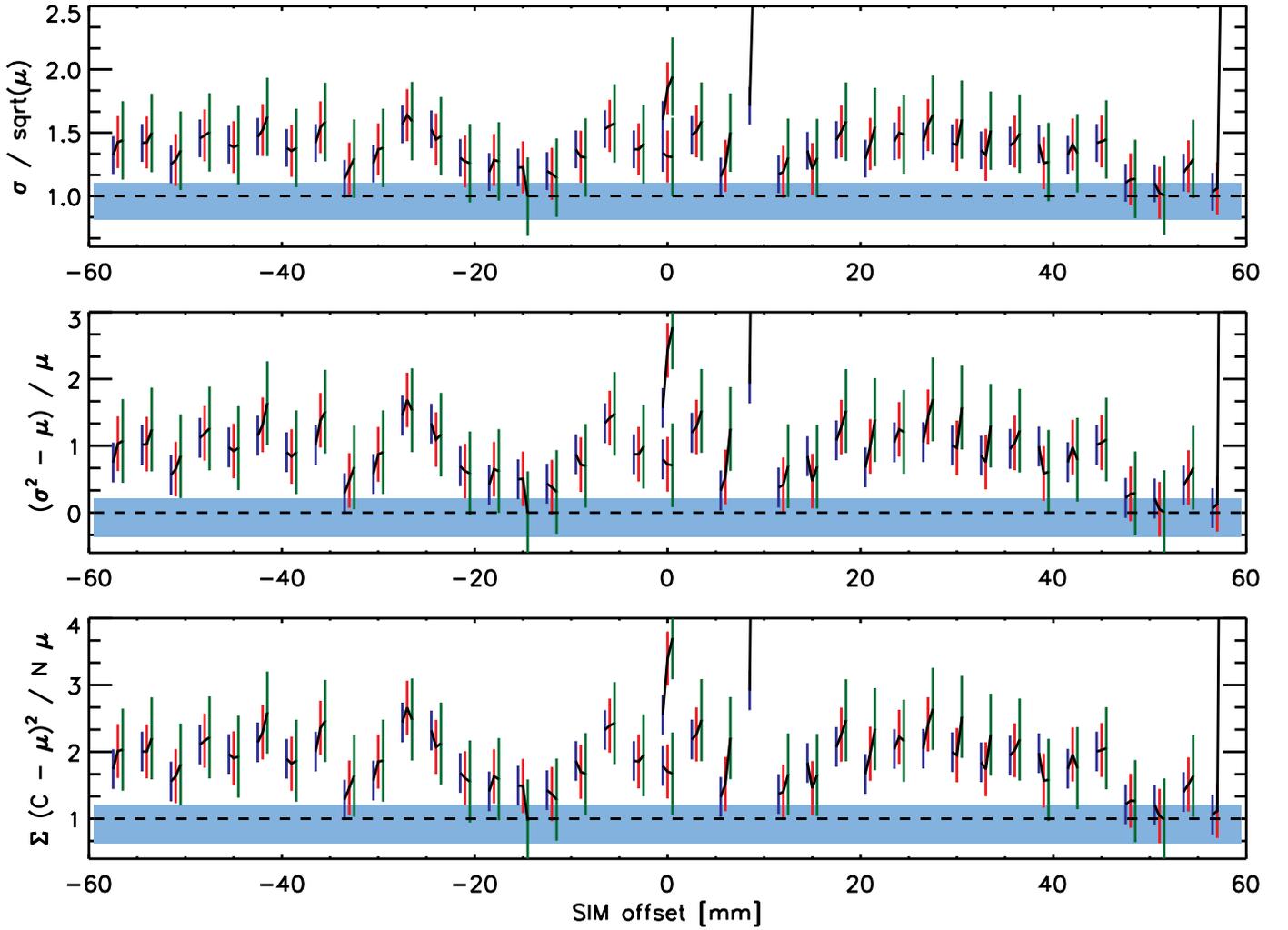}}	
\caption{Overdispersion in the light curve in each Capella ObsID.
The overdispersion measures $\rfrac$ (top), $\rdiff$ (middle), and
$\rchis$ (bottom) are calculated for each ObsID for different
values of the light curve bin sizes $\dt=25,50,100$~s, and
are denoted by the thin vertical lines grouped around the SIM
offset for that observation.  The lines are offset from each
other for clarity and have $\dt$ increasing from left to right,
and the measured values for each ObsID are connected by dark lines.
The vertical lines represent the $\pm3\sigma$ error bars for the
{\sl null}, determined from Monte Carlo simulations of a model
without any intrinsic variability but matching the count rate and
exposure time of the observation.
The values expected for the null model are shown for each
$\rho_{(\cdot)}$ as the horizontal dashed line.
The overdispersion measures computed for the combined \hz43\
data (for a binning that matches $\dt=25$~s for Capella) is
shown as the pale blue band whose width corresponds to the
$\pm3\sigma$ error bounds determined the same way as for
Capella.
\label{f:overdisp}}
\end{figure}

The \hrci\ light curve (Figure~\ref{f:ctrt}) shows numerous sharp
fluctuations similar to that expected from Poisson fluctuations.
We have tested whether these fluctuations are consistent with such
a picture, and conclude that they are not; the observed fluctions are
invariably {\sl overdispersed} compared to the expected Poisson
deviations (see Figure~\ref{f:overdisp}).
In particular, given a counts light curve $\lc(t_i;J,\dt)$ for ObsID $J$,
where $t_i$ are the $N_\dt$ time bins resulting from choosing a bin of
size $\dt$, we summarize the light curve with its mean
$
\mu_J(\dt) 
$
and variance
$
\sigma_J^2(\dt) 
$
and compute three measures of overdispersion:
\begin{mathletters}
\begin{eqnarray}
\rfrac &\equiv& \frac{\sigma_J(\dt)}{\sqrt{\mu_J(\dt)}} \,, \\
\rdiff &\equiv& \frac{\sigma_J^2(\dt)-\mu_J(\dt)}{\mu_J(\dt)} \,, \\
\rchis &\equiv& \sum_{i=1}^{N_\dt} \frac{[\lc(t_i;J,\dt)^2-\mu_J(\dt)]^2}{N_\dt\mu_J(\dt)} \,.
\end{eqnarray}
\end{mathletters}
If the light curve fluctuations are fully explained as Poisson
fluctuations, we must have
$\rfrac\approx1$, $\rdiff\approx0$, and $\rchis\approx1$,
since the variance of a Poisson process is equal to its mean.\footnote{
Note that unlike the autocorrelation analysis above, here we use
counts light curves in order to maintain the correspondence
with a Poisson statistical process; the effect of QE non-uniformity
is negligible, as described in \S\ref{s:data}, and as demonstrated
explicitly in \S\ref{s:hz43}.
}
In contrast, if an additional process is operating to cause
variations in the source intensity, we must have
$\rfrac>1$, $\rdiff>0$, and $\rchis>1$.

\begin{table}[htb!]
\begin{center}
\caption{Average overdispersion \label{t:overdisp}}
\medskip
\begin{tabular}{lccc}
\hline\hline
$\dt$~[s] & $\rfrac$ & $\rdiff$ & $\rchis$ \\
\hline
25 & 1.35$\pm$0.15 & 0.84$\pm$0.40 & 1.83$\pm$0.40 \\
50 & 1.41$\pm$0.31 & 1.08$\pm$1.26 & 2.06$\pm$1.25 \\
100 & 1.50$\pm$0.43 & 1.44$\pm$1.87 & 2.39$\pm$1.84 \\
\hline
\end{tabular}
\end{center}
\end{table}

For the \hrci\ dataset, computing the overdispersion measures
for all ObsIDs, for a variety of time binning sizes
($\dt=25,50,100$~s), we find that the latter condition holds
(Figure~\ref{f:overdisp}; see also Table~\ref{t:overdisp})
for the majority of the cases.
In each case, we compute the error bars on $\rho_{(\cdot)}$ via Monte Carlo
simulations of the null model, and thereby determine the significance
of each measurement of the overdispersion measure by calibrating the
statistic independently for the specific values of $\{\mu,N_\dt\}$.
Thus, we find that in most of the observations, the overdispersion is
significant, and hence that the observed variations in the
Capella light curves cannot be explained as due only to Poisson
fluctuations.
In particular, we find
$\rfrac\approx1.37-1.45$,
$\rdiff\approx0.87-1.09$, and
$\rchis\approx1.86-2.05$
as the range of the medians over the entire dataset for the
three different bin sizes considered.  The overdispersions
tend to increase with $\delta$, as is expected because the
relative statistical error decreases with increasing counts.
The averages and the standard deviations are reported
in Table~\ref{t:overdisp} for each $\dt$, and show that the
result is robust over different methods and binning sizes.
The measured points are in every case except one greater
than the nominal value for the null, and the large standard
deviations arise from large deviations upward.
Thus we conclude that Capella exhibits variability over
durations that characterize the observations, i.e., at
timescales $\sim5$~ks.  More sophisticated analyses (which
are in progress) are necessary to fully characterize this
variability.

These measurements suggest that the intrinsic variability has
an effect that is of similar magnitude to the Poisson process,
and that we are able to detect it primarily because of the
large count rates and because the large number of independent
datasets allows us to check that all give similar results.
Our analysis does not allow a direct calculation of the intensity
variations that result in these overdispersion values, but
assuming a model with small deviations, we estimate that
intensity variations of the order
$\frac{\Delta\lc}{\lc}\approx0.02-0.07$ are
consistent with the observed $\rho_{(\cdot)}$.
Since such intensity variations are of the same order as the
expected Poisson fluctuations in the light curves, it is not
surprising that methods such as the Kolmogorov-Smirnoff test
and Fourier analysis fail to detect the existence of the subtle
variability in the data.
Similarly, Argiroffi et al.\ (2003) place a $3\sigma$ limit
on the variability at $<5-10\%$, based on \chandra\ HRC-S/LETGS
data where Capella has a count rate of $\approx2.7$~\ctrt.
With the \hrci, count rates $\approx8$ times higher are observed,
which allows us to detect the existence of such variability
at $>3\sigma$ (Figure~\ref{f:overdisp}).

\subsection{Calibration \label{s:hz43}}

As the source position changes on the detector with each observation,
and even during a single observation as it dithers across the
detector, the source passes over regions with different quantum
efficiencies (QE).  Variations in the QE will lead to corresponding
differences in the observed count rate, and may produce a false
signal of variability.  In comparing the source intensities from
different ObsIDs, we have taken the QE variations into account
using the current best estimate of the \hrci\ QE map, which is
estimated to be accurate to $\approx5\%$ across the detector
(Posson-Brown \& Donnelly 2004; Figure~\ref{f:ctrt} suggests that
the relative error in the QE map is considerably lower).
While computing the overdispersion measures however, we do not
make such corrections in order to not bias the statistical
estimates.  Instead, we estimate the contribution of the local
variations in QE to the overdispersion by comparing the variances
in the light curves generated with and without correcting for
this effect: the
observed $\sigma(\dt)$ increases by $\lesssim3\%$, which is
negligible compared to the magnitude of the observed overdispersions
(Figure~\ref{f:overdisp}).
As a further test, we have carried out Fourier transforms of
the counts light curves; these show no evidence of a periodic
signal, as would be present at frequencies corresponding to
the dither periods if QE variations were to make a significant
contribution to the variability.
In addition, we place a direct limit on the residual calibration
uncertainty by comparing the analysis of Capella with that of
a known non-varying source, \hz43.


\hz43\ is a H-rich DA white dwarf with a temperature
$T_{\rm eff}\approx50$~kK (Dupuis et al.\ 1998).  It has no known
or expected variability, and thus serves as a comparison target to
verify the effect of the residual calibration uncertainty.  The
source has been observed numerous times as a calibration target
by \chandra\ in the \hrci+LETGS configuration, and has thus far
been observed for a total of 31.3~ks at the nominal aimpoint.
The observed count rate is 3.6~\ctrt, and coincidentally the
accumulated counts
approximately matches the number of counts produced in any given
Capella ObsID, making it an excellent proxy to test for the
existence of any instrument-based variability.\footnote{
\hz43\ has also been observed at other times and other locations
on the \hrci, but these observations do not have sufficient counts
for a useful comparison with Capella data.  Furthermore, the
off-axis pointings also have large PSFs, which again precludes
direct comparisons.
}
We measure
the overdispersion in \hz43\ at various bin sizes, set such
that the same number of counts are expected in each bin as
for the Capella data.  For instance the Capella light curve
binned at $\dt=25$~s is approximately matched by a \hz43\ light
curve binned at $\dt=140$~s; i.e., absent intrinsic variations,
and assuming that the magnitude of the local QE variations are
spatial scale independent, both curves should have identical
statistical properties.
We find that the \hz43\ overdispersion measures are fully
consistent with there being no intrinsic variability whatsoever
(see Figure~\ref{f:overdisp}), and thus serve to confirm
the detection of overdispersion in the Capella data.

%

\subsection{Coronal Structure}

Stellar coronae are generally considered to be analogous to the
solar corona, in the sense that the X-ray emission arises from
optically thin, collissionally excited plasma, which is organized
in active regions by magnetic fields and is probably heated by
magnetic reconnection events.  In the case of Capella, the
energetics indeed support this view: the surface area of both
components is $\approx100$ times that of the Sun (Table~\ref{t:capella}),
and the X-ray luminosity is correspondingly higher, at $\approx100$
times the solar luminosity at the peak of its activity cycle.  This
is consistent with the picture of a solar like atmosphere, with the
coronae dominated by loops in active regions that cover a large
fraction of the surface.  However, the temperatures on Capella's
coronae are significantly hotter, and the strong emission component
at $T\sim6$~MK has no corresponding structure on the Sun.

We have established above (\S\S\ref{s:autocorr},\ref{s:overdisp})
that Capella exhibits variability over timescales ranging from
$\tscale\sim5-20$~ks.
While the longer timescale variability could arise simply due
to slow evolution of active regions, the variability at shorter
timescales ($\gtrsim5$~ks) is likely due to the dynamical balance
of heating events and cooling.  Because Capella maintains its luminosity
to within a few percent, it must be that heating and cooling events
are in balance for the most part, and therefore the variability
timescale must be matched to the total coronal heating rate changes,
and hence to the radiative cooling timescales.
(For typical coronal densities, conductive cooling is not a factor
at these timescales.)
We can estimate the physical characteristics of the Capella coronae
based on this correspondence.  The cooling timescale,
\begin{equation}
\tau = \frac{k_{\rm B} T}{n_e \Lambda(T)} \,,
\end{equation}
where $n_e$ is the electron number density, and $\Lambda(T)$ is
the power emitted by a unit volume of the plasma at temperature $T$.
Adopting a coronal temperature of $\log{T}=6.8$, and a
metallicity $Z=0.6$ (see Brickhouse et al.\ 2000), the power
emitted is $\Lambda\approx2.3\times10^{-23}$~ergs~cm$^{3}$~s$^{-1}$,
and we estimate the coronal number density
\begin{equation}
n_e = 4 \times 10^{9}~\tau_{10}^{-1} ~~ {\rm cm^{-3}} \,,
\end{equation}
where $\tau_{10}=\frac{\tau}{(10~{\rm ks})}$.
This is consistent with the limit $n_e\lesssim10^{10.2}$~cm$^{-3}$
found by Ness et al.\ (2003) based on an analysis of the
Ne\,IX triplet in \chandra\ and XMM-{\sl Newton} grating data.
For the observed
luminosity (see \S\ref{s:data}) and the adopted temperature,
the volume emission measure $EM\approx10^{53}$~cm$^{-3}$.
Assuming that the emission is spread uniformly across the
surface of both stars, and adopting a visible surface area 100
times that of the Sun, we estimate the height of the corona
\begin{equation}
h_{\rm est}\approx1.3\times10^{9}~\tau_{10}^2~f_{\rm fill}^{-1} ~~{\rm cm} \,,
\end{equation}
where $f_{\rm fill}$ is the surface filling fraction of
active regions.
This is comparable to the typical loop sizes seen on the Sun,
but is small compared to the density scale height on Capella,
$h_{\rm scale}\approx8\times10^{11}$~cm.

We assume here that the entire corona is part of the balance
between the heating and cooling occuring at the detected
variability timescales.  It is possible however that
there are multiple components of emission in the corona: the
small timescale variability may be due to a small portion of
the corona, e.g., as in X-ray bright points on the Sun, and
the bulk of the emission may arise in a lower density component
that cools slowly and is heated correspondingly slower and
reaches to a height comparable to the density scale height.
However, this scenario appears to be ruled out by the analysis
of high-resolution EUV spectra by Sanz-Forcada, Brickhouse,
\& Dupree (2003), who suggest that the bulk of the plasma on
Capella is at high densities.  Note that while we have confined
our attention to $\tscale=5-20$~ks, we cannot formally rule out
variability at shorter timescales (more sophisticated analyses
are in progress).  Such variability will act to increase the
estimate for the coronal density $n_e$ and decrease the height
of the emission layer $h_{\rm est}$.  We speculate that the
most probable scenario for emission on Capella is one where it
is dominated by numerous low-lying activity sites.

%
%
%

\section{Summary \label{s:summary}}

We have analyzed a deep \chandra/\hrci\ observation of the
active binary Capella to detect variability at timescales
shorter than 50~ks.  Capella is a highly stable coronal
source on which hitherto no short term variability had been
seen, despite its relatively high activity level.
The \hrci\ is uniquely positioned to achieve this objective
because it allows the measurement of large count rates
($\gtrsim20$~\ctrt) with no pileup effects.  The unprecedented
data quality allows us to test for the existence of variability
at timescales as short as 5~ks.

We confirm the conclusion of Argiroffi et al.\ (2003),
Raasen et al.\ (2007), and Westbrook et al.\ (2007) that
Capella exhibits intensity variations at the $3-10\%$ level
over timescales of months and years, but unlike those studies
which were limited in statistical power due to low count
rates, we detect variability at timescales $<50$~ks.

We apply numerous statistical tests such as autocorrelation,
overdispersion, K-S, etc., to the data, both cumulatively and
in individual observation segments, and find that variability
does exist at timescales $\tau=5-20$~ks.  This suggests that
the coronal plasma is at a relatively low density
($n_e\lesssim10^{9}$~cm$^{-3}$) and that the emission arises
in low-lying loops with heights $\sim10^{9}$~cm.

\acknowledgements
This work was supported by the Chandra X-ray Center NASA contract
NAS8-39073.  We thank Brad Wargelin, Frank Primini, Jeremy Drake,
Mike Juda, Nancy Evans, and Scott Wolk for useful discussions.


%

\end{document}